\documentstyle[aps]{revtex}

\begin{document}
\draft
\title{Dynamical ordering induced by preferential
transitions in Planar Arrays of Superheated Superconducting
granules}
\author{A. Pe\~{n}aranda\footnote{angelina@fa.upc.es} and 
L.\ Ram\'{\i}rez-Piscina
}
\address{
Departament de F\'{\i}sica Aplicada,\\
Universitat Polit\`{e}cnica de Catalunya,\\
Avda. Gregorio Mara\~{n}on 44, E-08028 Barcelona, SPAIN.}
\maketitle

\date{\today}

\begin{abstract}
We perform simulations of planar arrays of superheated superconducting 
granules (PASS) under  an external magnetic field, analyzing transitions 
undergone by the system when the external field is slowly increased from zero. 
We observe, for high concentrations, the existence of an interval of external
fields for which no transitions are induced. This effect is analogous to a
 "hot border zone" identified in the response of
 Superheated Superconducting Granule detectors. We explain such behaviour as produced by
a geometrical ordering dynamically induced in the system by 
transitions in preferential sites due to diamagnetic interactions.

\end{abstract}

\thispagestyle{empty}
%\newpage

\vspace{1cm}

Superheated superconducting granules systems are being developed as detectors 
in areas such as dark matter, neutrino, neutron, x-ray  and transition radiation
 \cite{bern1}.
An ensemble of spherical granules of a Type I superconductor material is 
maintained in a
metastable superheated state by adequate conditions of temperature and external
 magnetic field. An increasing of the applied field or the deposition of energy 
in a microgranule by radiation can produce a transition 
to the normal state.  The change of magnetic flux inherent to the loss of 
Meissner effect can be sensed by a sensitive magnetometer which provides
information about the incident radiation.

An early detector was proposed as a disordered colloidal 
suspension of microgranules in
 a suitable medium such as paraffin wax \cite{mettout}. 
The state of each grain of this suspension 
in the phase diagram depends, in addition to the external field and temperature,
 on its metallurgical defects and the diamagnetic interactions of the other 
grains of the suspension. Metallurgical defects can increase the local 
magnetic field of the grain and can precipitate the transition to normal state.
 Diamagnetic interactions depend on the environment of each grain, producing
 an additional dispersion
  in the surface magnetic field values. As a consequence of these combined 
effects, the spread of transitions fields of the suspension reaches values of 
 about $20\%$ that can reduce the resolution of the detector \cite{drukier}.

In a previous work we showed that this spreading in transition fields is 
effectively reduced following an increase in the applied field.
 We obtained that the successive transitions induced by the external field 
are a strong ordering mechanism
 which produce a more homogeneus distribution of surface magnetic fields of 
the granules. Consequently, by using this effect, the uncertainty could be reduced 
in these devices \cite{Penaranda2,Penaranda3}. 

Experimentally, a variant of the colloidal device has been developed in
response to the spread
 problem \cite{PASS}. The microgranules are arrayed on thin planar substrates. 
These Planar Arrays of Superconducting Spheres (PASS) have yielded differential
 superheating curves in which the spreading  is reduced by an order of 
magnitude.

Although the technique of fabrication of this device can only produce arrays 
of relative small size, it has been shown that the PASS has both good 
energy and position sensitivity. On the other hand, the avalanche effect, 
demonstrated
 in lines of granules, can enhance the magnetic signal. This allows, in 
principle, the use of very small grains in devices with high 
energy sensitivity \cite{avalancha}.     
We presented in a previous work \cite{Penaranda1} simulations of
these systems, and results of maximun 
surface fields were shown for different distances between spheres, i.e. 
different concentrations. The broadening of maximum surface field 
distributions for increasing concentrations as a consequence of 
diamagnetic interactions, with noticeable finite size effects, was noted.
This would produce a larger dispersion in transition fields.

Numerical Monte Carlo techniques were applied by Esteve et al.
\cite{Esteve} to study transitions induced in a bidimensional
lattice of small superconducting spheres by an external magnetic field
normally applied to the array.
Even though they only considered two body interactions and diamagnetic
contributions until a small number of neighbours of each sphere,  
they found an interesting phenomenon: a  gap or plateau zone appears in 
the curves of transition counts versus the applied field for
systems with small lattice spacing, which seems always located around a
 fraction $f$ of remaining superconducting spheres equal to $0.30$. 
At this point, a large increase of field is required to observe the next
transition.
The qualitative explanation given for the existence of the plateau is
the decreasing of diamagnetic interactions due to the change of distance 
between spheres as a result of transitions. Nevertheless the authors did not provide any
 explanation of the observed discontinuity nor the $f=0.30$ occurrence 
of this phenomenon. 	

This plateau corresponds to an effective 'hot border zone' previously observed
in the response of Superheaed Superconducting Grains devices, in the
 sense that the system is insensitive to small changes in external
field, and transitions can only be produced by thermal nucleation
\cite{girard1}. This effect has interesting consequences in PASS operations,  because it introduces a threshold in the value of the energy of the incident radiation to permit transitions to occur.

In this paper we  perform numerical simulations of transitions in PASS 
configurations immersed in an increasing external field, $B_{ext}$,
in order to analyse this phenomenon. We observe that the existence or not of 
the plateau is related to different 
dynamical evolutions, for concentrated or dilute systems, during transitions. 
This dynamics produces, in the case of more concentrated systems, a spatial order  
coherent with the appearance of the plateau zone at about $f=0.25$ value. 
This ordering is the result of previous preferential transitions in certain 
lattice sites due to diamagnetic interactions. However, 
transitions in dilute systems give an evolution to a different kind of 
spatial configuration. 
A key point in this differentiation is the role of diamagnetic interactions,
which are stronger in more concentrated systems. 

We consider the applied magnetic field $B_{ext}$ perpendicular to 
the planar system.
We assume the microgranules as spheres of equal radius $a$ 
much larger than the London penetration length. We consider that the 
transition of each microgranule to the normal phase is completed once 
the local magnetic field at any
point of its surface reaches a threshold value $B_{th}$. This value can
vary from sphere to sphere and is introduced in order to take into account 
the defects of the spheres.

We employ, in our simulations, 
a distribution of threshold values experimentaly determined for tin 
microspheres dispersed in paraffin. 
This distribution was fitted  by a parabolic distribution
in a range of values between $B_{SH}(1-\Delta)$ and $B_{SH}$
\cite{geigenmuller}. Small values of $\Delta$ are related
to more perfect spheres. Most results shown in this work correspond to 
$\Delta=0.2$.
  
The procedure in our simulations is as follows: $N$ spheres are placed 
in a square array, separated a distance $d$.
 After assignment at random of a  threshold field value 
 to each sphere, the system is immersed in an external magnetic 
field $B_{ext}$  which is slowly increased from zero.
The knowledge of the surface magnetic field on each microsphere is achieved 
by solving the Laplace equation with the appropriate boundary conditions.
We have used a numerical procedure that allows us both
to consider the complete multi-body problem and to reach multipolar
contributions of arbitrary order \cite{Penaranda1}.
When the maximum local magnetic field on the
surface of any sphere reaches its threshold value $B_{th}$,
the sphere transits and
the configuration becomes one of $N-1$ superconducting spheres.
The change of diamagnetic interactions in any transition leads
us to repeat the process until all spheres have transited. 
The maximun surface magnetic field value of each sphere is monitored after each 
transition, allowing us to study the evolution of a system in its successive 
transitions.

Numerical simulations have been performed on 
several configurations with distances between sphere centers, 
in units of radius $a$, of $d/a= 7.482, 4.376, 3.473, 3.034, 2.757$ and $ 2.5$.
These distances correspond in a 3-D array to values of 
filling factor (fraction of volume occupied by the spheres), of 
 $\rho= 0.01, 0.05, 0.10, 0.15, 0.20$ and
$0.268$. 

An important point to be considered in this kind of system is the finite-size 
effects \cite{Penaranda1} that affect surface magnetic field values, 
especially in dense configurations, and that force us to work with a 
number of microgranules as  large as possible \cite{Penaranda1}. 
By computational limits and precision requirements, 
the number of spheres analysed has been $N=400$ in the 
more concentrated systems and $N=169$ in the other configurations.

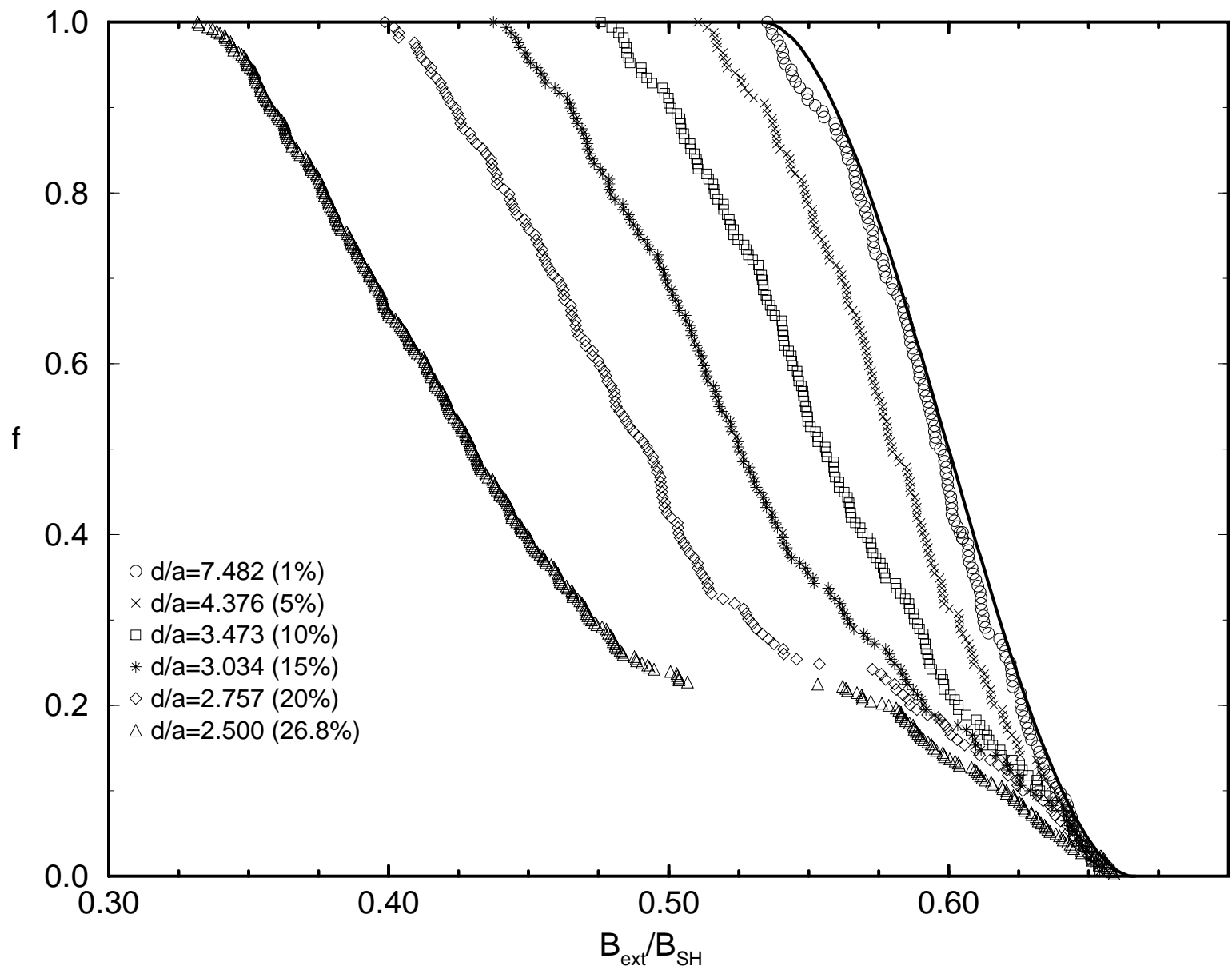
\begin{figure}
\vspace{30mm} % height of figure
\caption{
Fraction  $f$ of still superconducting spheres versus $B_{ext}/B_{sh}$,
after
an increase of the perpendicular external magnetic field from zero,
for several samples of
$N=169$ initially superconducting spheres,
corresponding to different initial lattice spacings. ($N=400$ for 
the more concentrated systems).
Continuous line corresponds to the dilute limit, {\it
i.e.} assuming a maximum surface field of $3/2 B_{ext}$ for all the
spheres.
}
\label{fbper}
\end{figure}

Results of simulations of field-induced transitions are shown
in Fig. \ref{fbper}. In this figure, the fraction $f$
of still superconducting spheres versus
the (increasing) external field, refered to the critical superheating field
($B_{ext}/B_{SH}$), is presented for several values of lattice spacing.
We can observe a fast decay in the most dilute case, in which transitions are 
produced for external field values closely following the distribution
of threshold values.
Transitions begin for smaller external fields values as the concentration of
the system increases. This shows the significance of diamagnetic
interactions on local surface field values, which is stronger for spheres
 in closer proximity. 
 On the other hand, the transition curves spread out for these concentrated 
configurations. 
But the more significant effect shown in this figure is the breakdown of 
the continuous response and the appearance of a  
 'plateau zone' clearly distinguished for shorter lattice spacing.
 This effect is 
produced for a fraction of remaining superconducting spheres slightly
lower than $f=0.25$.
In this zone there is a gap in the necessary increment of the external field 
to generate the following transition. The width of this gap increases as the lattice
spacing is reduced.
This plateau corresponds to an effective 'hot border'.

Comparison with results from the work of Esteve et al. \cite{Esteve}, shows  
 great similarity even though the location of the plateau is in their case always
 around $f=0.3$. They worked with perfect spheres and two-body interactions, which were only considered for spheres closer than a few lattice spacing.
 They interpreted this zone as an interpolation
between two qualitatively different dilute regimes. One would correspond to a  
initially homogeneous system, for large  values of $d/a$, where diamagnetic 
interactions are not very important. The other corresponds to a  regular 
configuration obtained as a consequence of the dilution of an initially 
concentrated system after transitions.

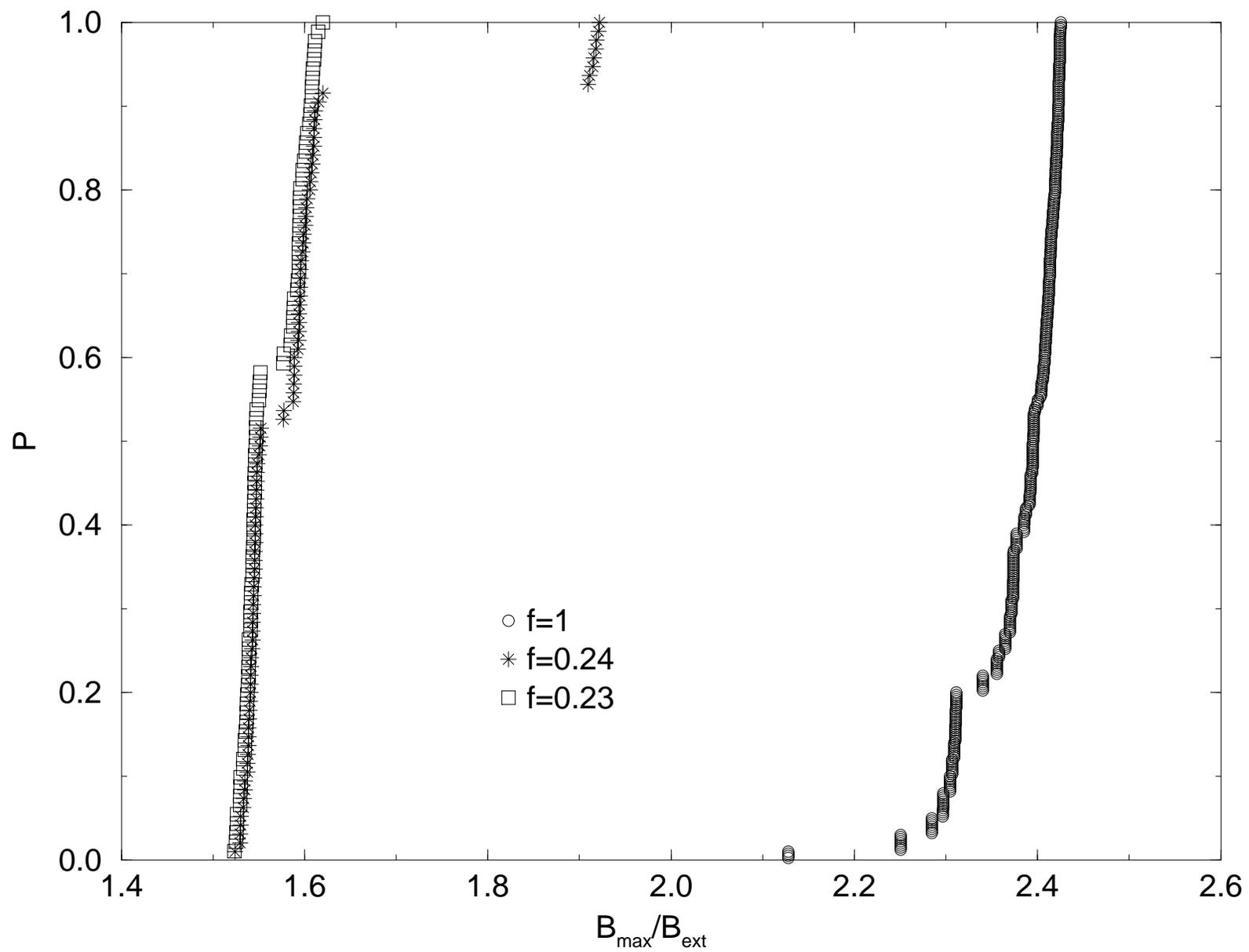
\begin{figure}
\vspace{30mm}
\caption{
Fraction $P$ of spheres with maximum surface field lower than the
$x$-axes value (in units of $B_{ext}$), in the evolution of a
configuration with initial lattice spacing $d/a=2.5$ ($\rho=0.268$)
and $N=400$ near the plateau zone. 
}
\label{figevoperb}
\end{figure}

We analyse this effect by studying the dynamics of the system
in its evolution during the increase of the external field. We consider both
the spatial
configurations and the distributions of surface fields that change after each
transition. 
Some of our results are represented in Fig. \ref{figevoperb}, where the maximum surface magnetic 
field distributions for a configuration with $d/a=2.5$ are shown  
at three values of the increasing
external field. Namely we present distributions for the initial state ($f=0$,
all $400$ granules are still superconducting) and for configurations before 
and after the plateau ($f=0.24$ and $f=0.225$, $96$ and $90$ superconducting
 spheres respectively).
Transitions induced by the external magnetic field split the initial distribution 
 in two branches separated by a gap. When the system is reaching  
the plateau zone, only a small number of spheres are in the branch of 
high surface magnetic fields.
Some of these microgranules will be the next to transit.
Each transition affects the interactions between microgranules, especially in 
the nearest neighbours by reducing their surface field. 
This situation is reflected 
in Fig. \ref{figevoperb} by the jump of each sphere from one branch to the
other. The disappearance of the high field branch corresponds
to the plateau zone. The remaining 
superconducting spheres have lower maximum surface fields,
and  need larger external fields to achieve their
threshold value and turn to the normal state. This explains the presence of 
the plateau. 

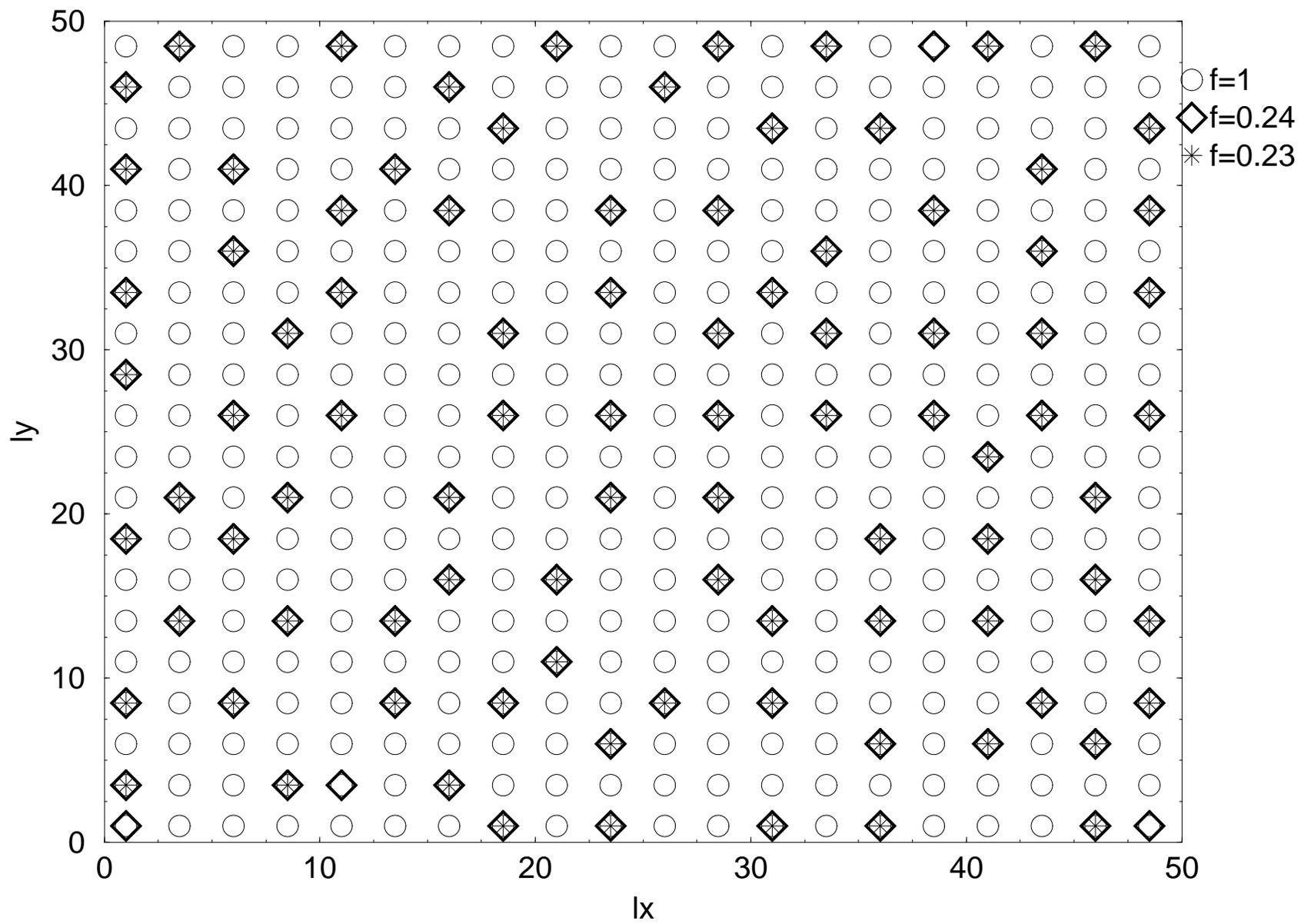
\begin{figure}
\vspace{30mm}
\caption{
Spatial distribution of initial $N=400$ spheres with lattice distance $d/a=2.5$
 ($\circ$), and the still superconducting spheres just before 
 ($\diamond$) and after ($\star$) the plateau zone.
}
\label{fotoperb}
\end{figure}

Looking closely at the spatial distributions, we observe that 
the branch with larger surface magnetic fields corresponds to spheres having
a superconducting next neighbour, and hence experiencing stronger diamagnetic
interactions.  In the plateau zone, only spheres without superconducting 
next neighbours remain superconducting. The system reaches a quite regular 
configuration with a fraction of superconducting spheres of about $f=0.25$. 
This is clearly
shown in Fig. \ref{fotoperb}. In this figure a snapshot of positions of 
superconducting microgranules are represented 
just before and after the plateau zone.

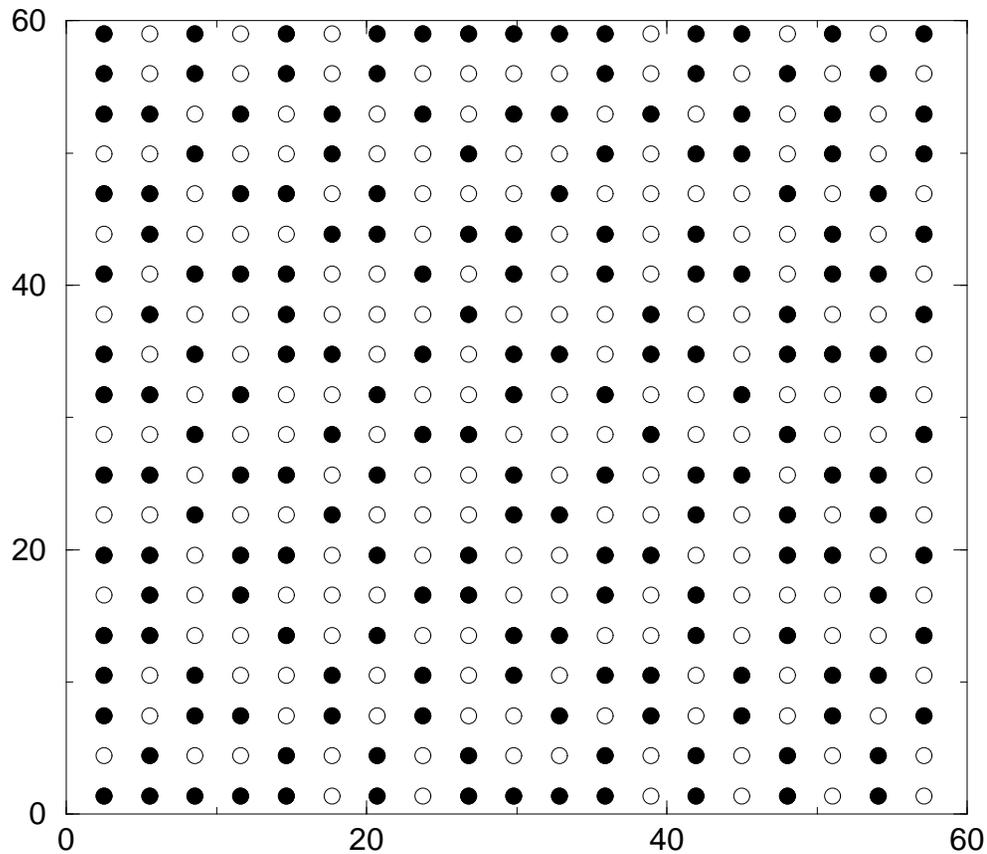
\begin{figure}
\vspace{30mm}
\caption{
Spatial distribution of initial $N=400$ spheres with lattice distance 
$d/a=3.034$
and $2.50$ (3D $15 \%$ and $26.8 \%$ respectively) and the corresponding distribution
when half of the microspheres have transited ($f=0.5$). 
}
\label{spatcomp}
\end{figure}

An interesting question is how the dynamics of transitions leads the 
system to such ordered configurations, and why this occurs for higher 
concentrations and not for dilute systems.
In order to gain insight into this phenomenon, we have studied in detail the
dynamics of transitions in systems with a reduced dispersion of threshold
fields (i.e. more perfect granules) and a larger number of initially 
superconducting spheres (in order to reduce finite-size and boundary effects).
These systems show a more perfect spatial ordering at the plateau (with 
$f$ very close to $0.25$) and therefore are more suitable to analyse regarding spatial configurations during transitions. This study reveals an
 interesting behaviour. These more perfect spheres present two clearly
different spatial distributions before reaching the plateau 
depending on the concentration (and 
corresponding to the the appearance or not of the plateau at $f=0.25$).
This is shown in Fig. \ref{spatcomp}
where the two different behaviours are compared at $f=0.5$ on systems 
with initial spheres separated distances $d/a= 2.5$ and $3.034$ 
(3D filling factor $\rho=20\%$ and $15\%$). 
For smaller distances between spheres (more concentrated systems, 
Fig. \ref{spatcomp}.b), the remaining superconductor spheres, after a number 
of transitions, show a configuration separated into domains. In each of 
these domains, transitions are produced in such a way that spheres have a 
tendency to form parallel lines in a sort of 'striped' configuration.
Until $f=0.5$ only spheres between lines transit. The resulting configuration is formed  by alternately superconducting and normal lines. 
For more diluted systems (Fig. \ref{spatcomp}.a) this
patterning does not exist.
Subsequent transitions in the concentrated systems are produced in such a way that 
 in each line, transitions occur of granules with superconducting
next neighbours. When the plateau appears, only spheres with third 
neigbours remain superconducting, forming a square lattice of spacing $2d$. 
This corresponds to a value of $f=0.25$ for the appearance of the plateau.
In the systems with less perfect granules, the domains are smaller and
not so clearly defined, and the plateau can appear for slightly smaller
values of $f$ due to more important boundary effects between domains.

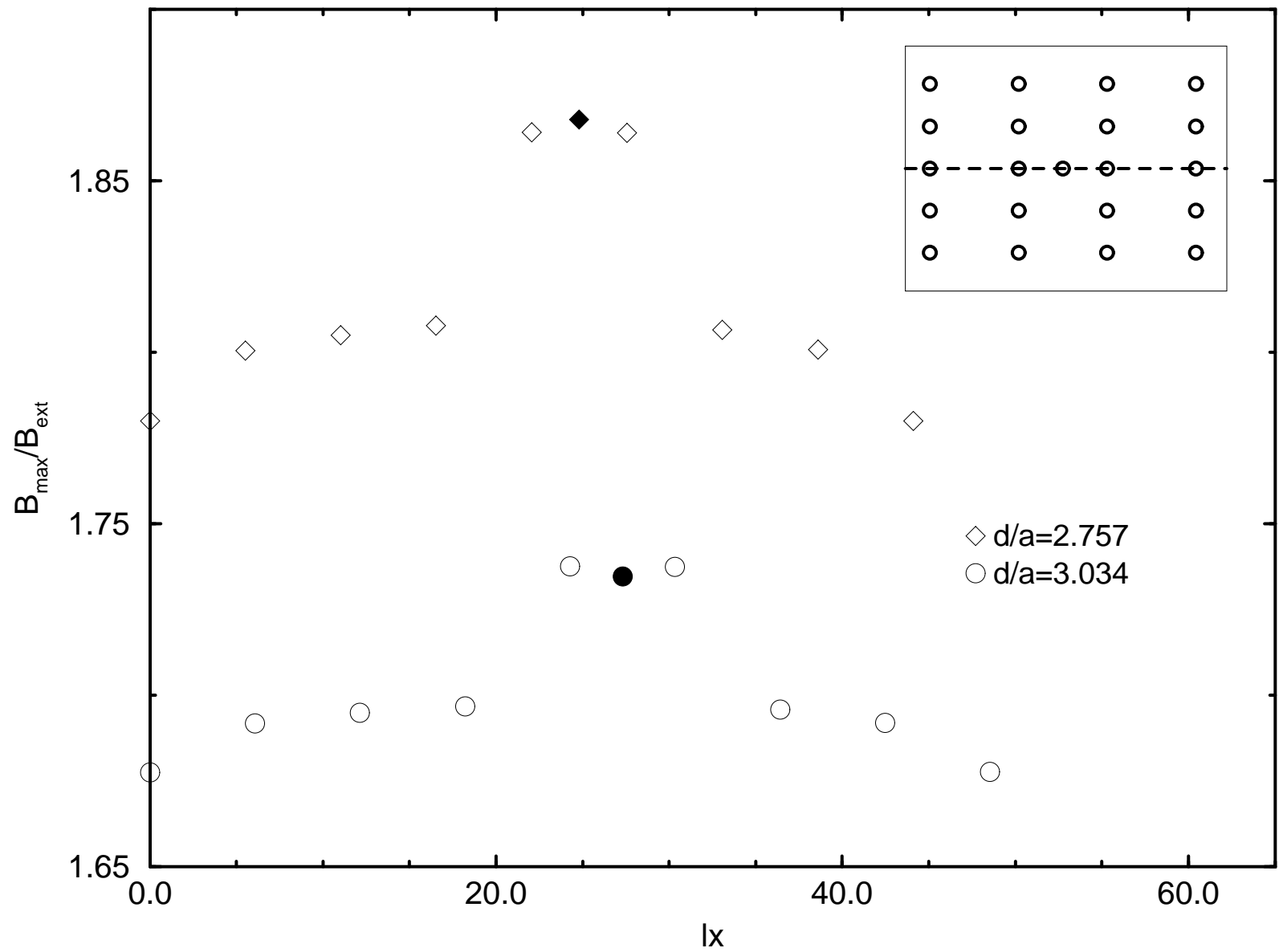
\begin{figure}
\vspace{30mm}
\caption{
Maximum surface magnetic field (in units of $B_{ext}$) for spheres 
, in a striped domain, with
 spatial configuration represented in the inset of the figure.
 The field values on spheres
lying on a line containing the central sphere are 
represented for $d/a= 3.034$ and $2.757$ ($\rho=0.15$ and $0.20$). 
}
\label{bmxcomp}
\end{figure}

From this dynamical study we observe that the ordered spatial configurations 
 for $f=0.5$, would condition the  existence or not of the plateau at $f=0.25$.
We have elaborate a criterion that allows one to know if the striped 
configuration is possible, at $f=0.5$, for a particular system.
This criterion uses the simulation of a system representing one of these
domains. We prepare this system with parallel stripes of superconducting 
granules and an additional granule in the middle.
This additional granule should be the first to transit in order to reach the striped
configuration. If it should not be so, this configuration could not
be possible and them the plateau would not appear.
We have performed  simulations on one of these domains by placing $N=82$
 spheres distributed in $9$ lines of $9$ spheres each, and the additional sphere 
 in a central position between two lines.
 Each line is separated a distance $2 d$ from the other. 
The distance between spheres of the same line is $d$. A diagram of this 
system is shown in the inset of Fig. \ref{bmxcomp}.
 Analysing the maximum surface fields of the spheres in this 
configuration, for different lattice values, 
we have observed that for diluted systems, the sphere that has the highest 
maximun surface field (and that will be the next sphere to transit) is not 
the additional one. On the contrary, for more concentrated
systems, the highest maximum surface field
does correspond to the central sphere, and consequently 
this sphere will be the first to pass to the normal state. In this case,  
the remaining spatial configuration will be formed by complete lines. 
This is displayed in Fig. \ref{bmxcomp} where the
maximun surface fields of the spheres on
the horizontal line containing the central sphere are presented 
for two representative values of $d/a$. 

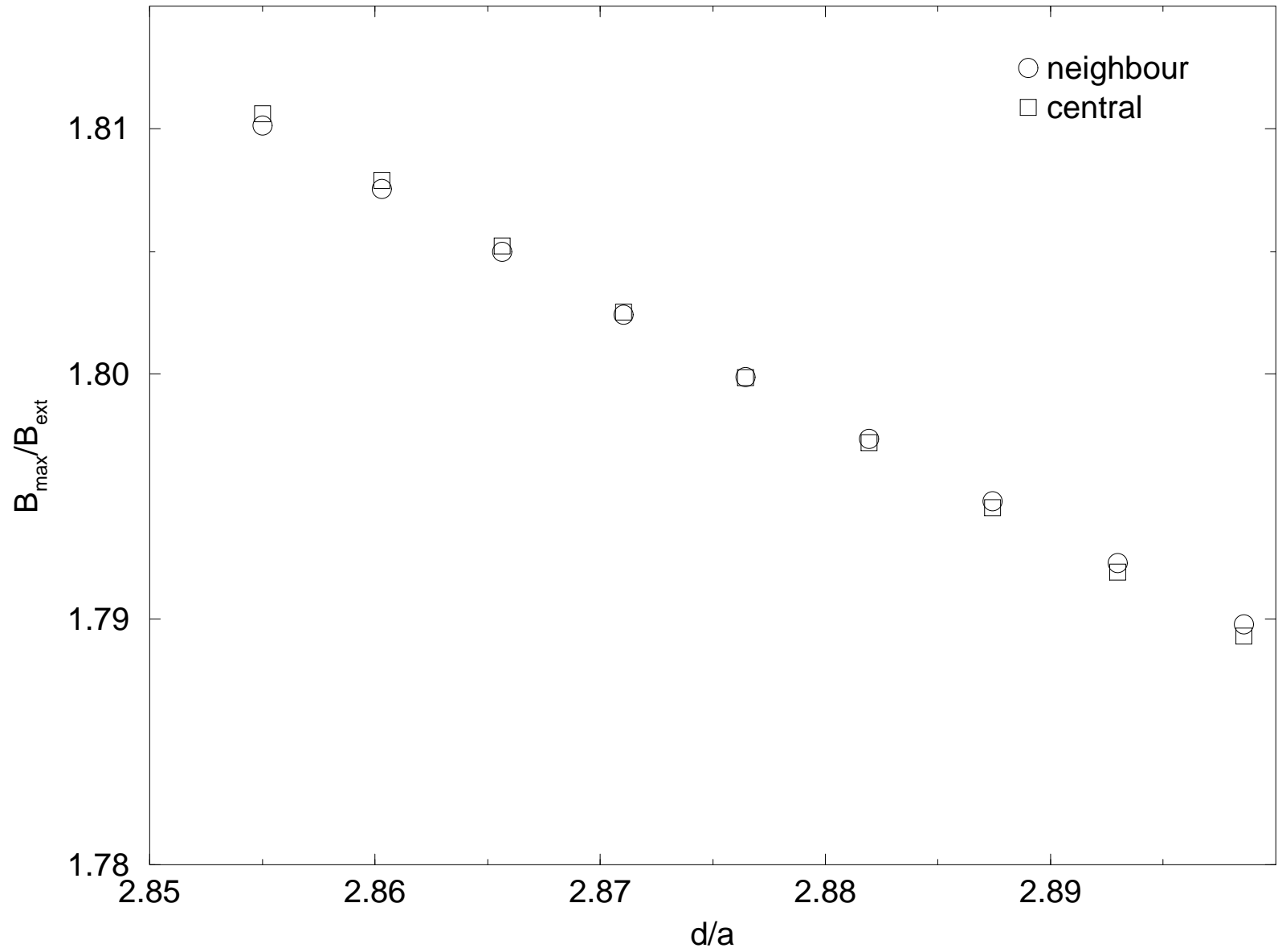
\begin{figure}
\vspace{30mm}
\caption{
Maximum surface magnetic field (in units of $B_{ext}$), versus lattice 
distances,
 corresponding to the central sphere and its neighbour of a striped domain . 
}
\label{denslim}
\end{figure}

Repeating simulations for different values of the lattice distances permits 
a location of the limit between both behaviours, and consequently 
the density above which the plateau zone appears. We have obtained this limit 
for a lattice distance $d/a=2.871$ ($\rho=17.7\%$)  in these ideal
 conditions. Results are represented in  Fig. \ref{denslim} for 
lattice distances near to this concentration. In this figure the maximum 
surface field of the central sphere and that of its next neigbour are compared.

It can be interesting to relate the response of the system to the applied 
field with  the position of the ensemble of spheres in the phase diagram. In both 
concentrated and dilute systems, the first 
sphere to transit will be that with largest diamagnetic interactions and 
consequently  with the highest maximum surface field (related to 
their threshold limit). In dilute systems, diamagnetic interactions are weak,
the maximum surface field values of the spheres have a small dispersion  
and the population of still-superconducting spheres will present a quasi-
continuous distribution in the phase diagram. Small changes of applied 
field can produce subsequent flips to the normal state, and the 
transition curves present a continuous aspect. For more
concentrated systems, the effect of diamagnetic 
interactions is very important. After successive transitions, the pairs of
nearest superconducting  spheres have higher surface fields, in comparison 
with those without superconducting next neigbours,
 as can be seen in  Fig. \ref{figevoperb}. This effect separates the 
population of still superconducting spheres in two distinguished groups 
in the phase diagram,  one, corresponding to spheres with higher surface fields
and near the superheated line, and separated from the other corresponding 
to spheres with smaller values. Successive transitions 
change the population of each group.
When the plateau appears, only the group of smaller field values remains superconducting. A small 
increment of the external field is unable to produce any transition; this is 
possible only by thermal nucleation. A larger 
increment of $B_{ext}$ is necessary
to continue the transitions. This is reflected as a 'gap', i.e. a plateau zone.   

We can conclude that diamagnetic interactions play an important role
in these kinds of systems, inducing distinct behaviours
depending on their concentration. In the case of small lattice
distances, a gap or plateau zone appears in the transition curves for a fraction of the remaining
superconducting spheres of
about $f=0.25$. This plateau is a consequence of a spatial order 
achieved through preferential transitions in these concentrated 
configurations. This order produces a uniform 
distribution of surface magnetic field values,  which is reflected in the
phase diagram as a distribution of the population of still superconducting spheres
separated from the superheated line. In this zone only transitions by finite
 increments of temperature are possible. This corresponds to a hot border.
Transitions undergone by dilute systems follow
different spatial distributions that do not bring the plateau appearance.
From simulation of ideal systems of quasi-perfect spheres, we have located 
the limit between the two behaviours at a lattice distance of $d/a=2.871$.

Finally, it is worth  to note the interest that this plateau has for a 
PASS detector, because the uncertainty in the
energy threshold for transitions can be reduced in 
the presence of a hot border.

\acknowledgements
We acknowledge T. Girard for helpful discussions and a careful criticism of the manuscript..
We acknowledge financial support from Direcci\'on General de
Investigaci\'on Cient\'{\i}fica y T\'ecnica (Spain)
(Project BFM2000-0624-C03-02, BFM2002-02629) and
Comissionat per a Universitats i Recerca (Spain)
Projects (2001SGR97-0021).
We also acknowledge computing support from Fundaci\'o Catalana per a la
Recerca-Centre de Supercomputaci\'o de Catalunya (Spain).

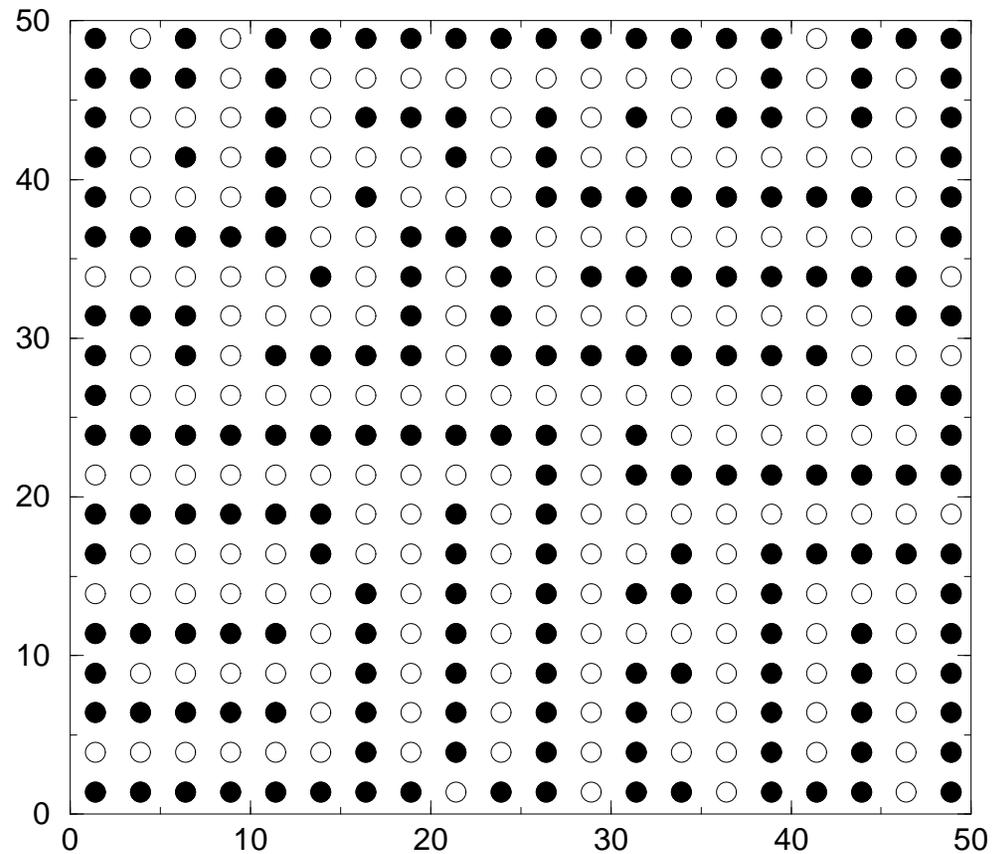
\end{document}